\documentclass[a4paper,fleqn,authoryear]{cas-sc}

\usepackage{lineno}

\nolinenumbers

\usepackage[numbers]{natbib}
\usepackage{float}
\usepackage{booktabs}

\def\tsc#1{\csdef{#1}{\textsc{\lowercase{#1}}\xspace}}
\tsc{WGM}
\tsc{QE}
\tsc{EP}
\tsc{PMS}
\tsc{BEC}
\tsc{DE}

\begin{document}
\let\WriteBookmarks\relax
\def\floatpagepagefraction{1}
\def\textpagefraction{.001}
\shorttitle{(Re)-Defining Planets}
\shortauthors{Madhu Kashyap Jagadeesh et~al.}

\title [mode = title]{Classification and Nomenclature of Planets in the Mass–Radius Plane}                      



\author[1]{Madhu Kashyap Jagadeesh}[type=editor,
                        auid=000,bioid=1,
                        role=Researcher,
                        orcid=0000-0003-4075-5646]
\cormark[1]
\fnmark[1]
\ead{madhu.kashyap@sju.edu.in}

\credit{Conceptualization of this study, Methodology, Software}

\affiliation[1]{organization={Department of Physics, St. Joseph's University},
                city={Bengaluru},
                postcode={560027}, 
                state={Karnataka},
                country={India}}

\author[2]{Arkil Parikh}

\author[3,4]{Margarita {Safonova}}[%
   role=Co-ordinator,
   orcid=0000-0003-4893-6150]

\credit{Data curation, Writing - Original draft preparation}

\affiliation[2]{organization={Christ (Deemed to be University)},
                city={Bengaluru},
                country={India}}


\affiliation[3]{organization={M.~P. Birla Institute of Fundamental Research},
                city={Bengaluru},
                postcode={560001}, 
                state={Bengaluru}, 
                country={India}}

\author[4,5,6]{Bernard Foing} [orcid=0000-0002-5110-1026]
\author[7]{Oleg Kotsyurbenko} [orcid=0000-0002-4748-3017]

\affiliation[4]{organization={LUNEX EuroMoonMars EuroSpaceHub},
                city={Leiden / Noordwijk},
                country={The Netherlands}}

\affiliation[5]{organization={ERA Chair on Space Photonics, University of Latvia},
                city={Riga},
                country={Latvia}}

\affiliation[6]{organization={COSPAR SCB Planetary Commission \& Panel on Exploration},
                country={International}}

\affiliation[7]{organization={NoRCEL Institute},
                city={Leeds},
                country={United Kingdom}}
\cortext[cor1]{Corresponding author}


\begin{abstract}
6500+ exoplanets have been detected using various techniques. This prompted the emergence of
many recent works on the taxonomy, or classification, of exoplanets. However, there is still no basic,
fundamental definition of `What is a planet?’. IAU has forwarded a definition in 2006, which
however, raised more questions than it solved. The first task here is to establish if there
are limits on the size/mass of planets. The lower mass limit may be assumed as of
Mimas (0.03 EU) – approximately minimum mass required to attain a nearly spherical hydrostatic
equilibrium shape. The upper mass limit may be easier – there is a natural lower limit to what
constitutes a star: ~0.08 SU. But then there are brown dwarfs: IAU has defined brown dwarfs as
objects exceeding the deuterium burning limit (~13 JU), and giant exoplanets generally have
masses of ~0.3 to ~60 JU. The resolution requires assembling the basic physical parameters that
define planets quantitatively. Mass and radius are the two fundamental properties, and we propose
to use a third correlated parameter: the moment of inertia. Based on this, we create the
parametric Fundamental Planetary Plane where the two parameters are correlated with the third. The fundamental planetary plane (FPP) with turn-off point diagrams is
constructed for visual representation. We propose an alternate potential description of a planet definition as ‘A celestial spherical object, bound
to a star or unbound, that lies on the fundamental planetary plane, within a mass range between
0.02 EU to 13 JU’. This definition is intended to complement existing taxonomies by providing a quantitative, structure-based criterion applicable to both Solar System planets, exoplanets and free-floating planets. These turn-off point diagrams serve as an alternative to the Hertzsprung–Russell (HR) diagram, but for planets.


\end{abstract}

\begin{graphicalabstract}
\includegraphics[width=1.1\linewidth]{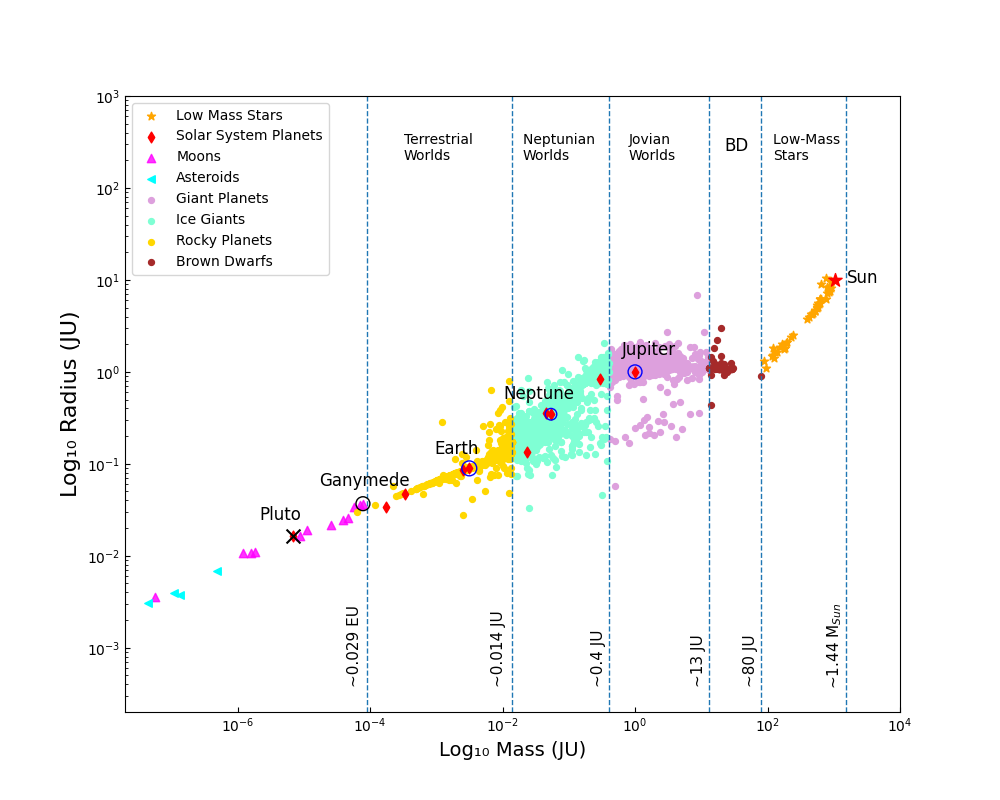}
\end{graphicalabstract}

\begin{highlights}
\item Introducing a ‘Fundamental Planetary Plane’, analogous to the HR diagram, that identifies planets using mass, radius, and moment of inertia, and demonstrates on physical grounds that Pluto does not occupy the planetary regime, independent of the neighborhood-clearing criterion.
\item Identifies a natural transition mass that distinguishes irregular minor bodies from gravitationally rounded planets, along with a corresponding trend in the radius–density relation.
\item Proposes a universal, physically motivated definition of planets applicable to both Solar System planets, free floating planets and exoplanets around the host star.
\end{highlights}


\maketitle

\section{Introduction}

What is a planet? The term `planet'  originates from the ancient Greek word {\it planētēs}, meaning `wanderer'. In modern age, Merriam-Webster encyclopedia\footnote{https://www.merriam-webster.com/dictionary/encyclopedia} defines a planet as "any of the large bodies that revolve around the Sun in the Solar System." Currently, 6500+ exoplanets are now detected using various techniques, with estimates of billions of planets existing in our Galaxy alone. They are called super-Earths, hot Earths, mini-Neptunes, hot Neptunes, sub-Neptunes, Saturns, Jupiters, hot Jupiters, Jovians, gas giants, ice giants, rocky, terran, subterran, superterran, and so on.

In 2006, the International Astronomical Union (IAU)\footnote{https://www.iau.org/Iau/Publications/List-of-Resolutions} passed a new resolution on the definition of a planet, which led to Pluto's reclassification as a dwarf planet. According to the IAU definition, we have three laws of planets:\\
{\bf Ist law of planets}: is in orbit around the Sun.\\
{\bf IInd law of planets}: has sufficient mass for its self-gravity to overcome rigid body forces so that it assumes a hydrostatic equilibrium (nearly round) shape.\\
{\bf IIIrd law of planets}: has cleared the neighbourhood around its orbit.

The IAU definition did not include exoplanets and free-floating planets. Also, the definition does not properly explain what it means to clear the neighborhood. Even the hydrostatic equilibrium clause is problematic -- for example, Mercury is not actually in hydrostatic equilibrium \citep{Balogh}, though nobody is suggesting to removing it from the family of planets.

These problems with the planetary laws stimulated a lot of work, and numerous new definitions and criteria have been proposed. In 2022, the IAU Working Group (WG) on Exoplanetary System \citep{2022L} suggested the following working definition of exoplanets: "Objects with true masses below the limiting mass for thermonuclear fusion of deuterium (currently calculated to be 13 Jupiter masses for objects of solar metallicity) that orbit stars, brown dwarfs or stellar remnants and that have a mass ratio with the central object below the L4/L5 instability", adding that it is irrelevant how they were formed. However, the free-floating bodies ('rogue planets') were still excluded from the planethood. 

More recently, \citet{2024M} proposed the following definition:\\
"A planet is that\\
(a) orbits one or more stars, brown dwarfs, or stellar remnants, \\
(b) is more massive than $10^{23}$ kg, \\
(c) is less massive than 13 Jupiter masses ($2.5 \times 10^{28}$ kg).\\
In addition, rogue planets ought to satisfy criteria (b) and (c) to be called a planet." 

"Clearing the neighbourhood", also called a dynamical dominance, means that a planet must be the gravitationally dominant object in its orbit. The idea was first introduced by Stern and Levison in a year 2000 to indicate that a planet must clear its orbit of the planetesimals during the final stages of formation \citep{2006S}. 

If a large object meets the other two planet criteria but hasn't cleared its neighbourhood, it is called a dwarf planet as is the case with Pluto, which shares its orbit with many Kuiper Belt objects and crosses paths with Neptune. 

Others have proposed ways to measure how well a body clears its orbit, using what they called planetary discriminants (\citealt{2006S, 2015M}). Some prefer the term dynamical dominance because it is clearer and less confusing (\citealt{2015M}). However, when it comes to planets outside our Solar System (exoplanets), it is still very hard to tell if they have cleared their orbits because we do not yet have the technology to detect small object near them (\citealt{2015M}). Even in our Solar System and only recently, using such powerful telescopes as HST and Keck Observatory, we started detecting objects we never seen before: such as, e.g., the hierarchical triple system discovered in not so far out Kuiper Belt (e.g. \citealt{altjira}). And even then it required accumulating data for 17 years. How can we expect to know the orbital neighbourhood of planets that are hundreds of parsecs away?

In addition, in our own Solar System, the planetary orbits are not completely clear. At least 100,000 Trojan asteroids near Jupiter's L4 and L5 points are still a mystery with respect to understand of their evolution and dynamics (\citealt{2023B}); it is obvious that Jupiter's neighborhood is not cleared. Earth immediate neighbourhood is very complex, it coorbits with many thousands of near-Earth-objects (NEO) with a discovery rate of over 3000 NEOs per year (\citealt{2023G}). According to recent studies, Earth temporarily captures mini-moons, which represent a subset of objects from the delayed bound population (\citealt{2025J}). It is possible that there are about 1,000 0.1-m diameter TCOs (temporarily-captured orbiters) in orbit around Earth at any given time \citep{Granvik}. Similarly, 20 co-orbital asteroids of Venus are known till date (\citealt{2025Ca}).
In addition, clarification is required regarding the precise meaning of the term dominance. For example, Earth is substantially more massive than the Moon and other bodies within its orbital region. However, if Earth were hypothetically located beyond Neptune, it would become gravitationally dominated by the icy giant. Under such circumstances, would Earth cease to satisfy the criterion for planetary status?



Several studies have previously explored quantitative approaches to planetary classification based on mass, radius, or density (e.g. Soter 2006; Margot 2015; Seager et al. 2007; Chen $\&$ Kipping 2017). In this work, we build upon these efforts by introducing a structurally motivated framework that incorporates the moment of inertia as an additional correlated parameter. While mass–radius relations capture bulk trends, the inclusion of the moment of inertia provides a complementary measure that reflects internal mass distribution and gravitational consolidation. The resulting Fundamental Planetary Plane is intended not merely as a visualization of known correlations, but as a physically motivated parameter space within which natural boundaries between planets and other classes of celestial bodies can be identified.

The structure of this paper is as follows. In Section 2, we present the mathematical framework underlying our approach, introducing the self-gravitating planetary model and the relevant scaling relations. Section 3 introduces the FPP and discusses its construction, including the identification of low- and high-mass turn-off thresholds that distinguish planets from other classes of celestial bodies. In Section 4, we examine the mass–radius correlation and assess whether the observed trends represent genuine physical relations or artefacts of modeling and data selection. An alternative diagnostic based on the radius–density plane is also explored to provide independent structural support for the proposed classification. Finally, we summarize our results and discuss their implications for planetary definition and taxonomy, along with directions for future work.

\section{Mathematical formulation}   

The mathematical formalism is done through a self-gravitating planet model with a \textit{polytropic equation of state}  
\begin{equation}
    P = K \rho^{1 + 1/n},
    \label{eq:polytropicEOS}
\end{equation}
where $n$ is a \textit{polytropic index}, determining the degree of central condensation \citep{1939C}. 
A polytrope is a model in which pressure $P$ and density $\rho$ follow this fixed power-law relation, allowing the description of its internal structure by the Lane--Emden equation,
\begin{equation}
 \frac{1}{\xi^2} \frac{d}{d\xi} \left( \xi^2 \frac{d\theta}{d\xi} \right) + \theta^n = 0   \,,
\end{equation}
where $\xi$ is a dimensionless radial coordinate, $\theta(\xi)$ is dimensionless function related to $\rho$ by $\rho=\rho_c\theta^n$, where $\rho_c$ is central density. From the Lane--Emden equation, the radius--mass ($R-M$) scaling is  
\begin{equation}
    R \propto K^{\frac{n}{3-n}} M^{\frac{1-n}{3-n}}\,.
\label{eq:radiusScaling}
\end{equation}
Here, $K$ is the real positive constant with dimensions that depend on the physical nature of the system (composition, entropy, etc.).\footnote{https://www.astro.princeton.edu/~gk/A403/polytrop.pdf}
\noindent
The moment of inertia is  
\begin{equation}
    I = k_n M R^2 ,
    \label{eq:momentInertia}
\end{equation}
where $k_n$ is a dimensionless constant determined by the polytropic index $n$. Uniform-density planets ($n=0$) have $k_n = 0.4$, while centrally condensed polytropes ($n>0$) have smaller values \citep{2007S}.

For most exoplanets and minor bodies, the moment of inertia cannot be measured directly and must therefore be inferred using simplified structural assumptions. In this work, we adopt a uniform solid-sphere approximation for all objects, including planets, moons, and asteroids. Although this approximation neglects detailed internal stratification, it provides a consistent baseline across a diverse population of bodies and allows comparative analysis within a common framework. Deviations from uniform density are expected to shift individual objects within the plane but are not expected to erase the large-scale clustering trends that define the FPP.


   \section{The Fundamental Planetary Plane} 

The Fundamental Planetary Plane (FPP) is introduced by direct analogy with the Hertzsprung--Russell (HR) diagram of stars. In the HR framework, luminosity is directly dependent on the stellar radius and effective temperature of the star through the Stefan--Boltzmann relation,
\begin{equation}
L \propto R^{2} T^{4}.
\end{equation}
The first step toward constructing such a framework for planets is to determine whether natural limits exist on planetary mass or size. A practical lower bound is often taken to be the mass of Mimas (approximately $0.03$~EU), representing the minimum mass required for attaining a nearly spherical, hydrostatic–equilibrium shape, consistent with the IAU second law of planet definition. However, exoplanet observations complicate this picture: the smallest confirmed exoplanet, Kepler-37b, has a mass of only about $0.01$~EU, as reported by standard exoplanet catalogues (e.g., Exoplanets Data Explorer\footnote{http://exoplanets.org}, Open Exoplanet Catalogue\footnote{https://www.openexoplanetcatalogue.com}). At the upper end, the conventional boundary between giant planets and stellar objects is set by the onset of deuterium burning at roughly $13$~JU, although known giant exoplanets span a broad range from about $0.3$ to $60$~JU. To resolve these ambiguities, we assemble the fundamental physical parameters that characterize planetary bodies—mass, radius, and a third structurally correlated quantity, the moment of inertia—and use them to construct the parametric planetary plane, in which any two parameters map uniquely onto the third.

\begin{figure}
\centering    
\includegraphics[width=1\linewidth]{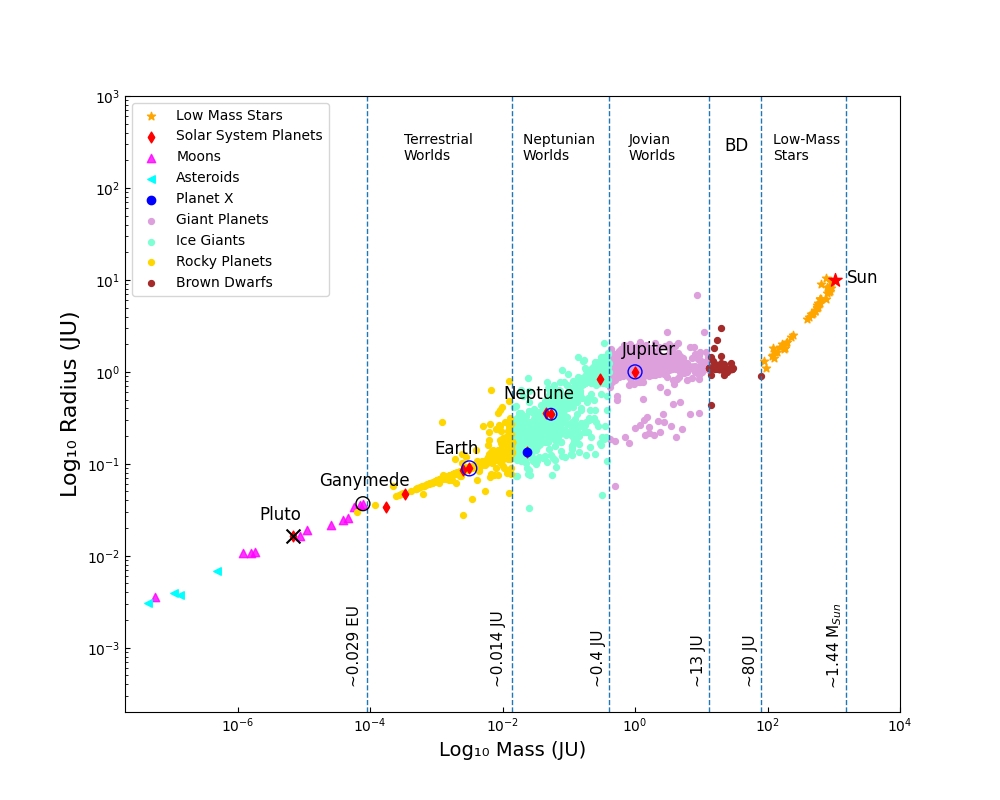}
\caption{Mass-Radius plot of asteroids, moons, planets, brown dwarfs and low-mass stars. From nearly 5600 exoplanets listed in the PHL-HWC database, 1504 are categorized as terrestrial worlds (orange dots), 2609 as Neptunian worlds (green dots), 1355 as Jovian worlds (purple dots), and 115 as brown dwarfs (brown dots).}
\label{MIU}
\end{figure}

In Fig.~\ref{MIU} we present the Fundamental Planetary Plane where we have plotted the entire range of celestial objects, such as asteroids, large moons and planets of the Solar System, exoplanets, and low mass stars. The data on exoplanets was taken from the PHL-HWC database, of Solar System objects and low-mass stars from the NASA Planetary fact sheet\footnote{https://nssdc.gsfc.nasa.gov/planetary/factsheet/} and \citet{2017C}, respectively. The exoplanets were classified into three groups (terrestrial, Neptunian, and Jovian worlds) based on the mass limits determined by \citet{2024Mu}. Objects having mass greater than the deuterium burning limit (\citealt{1997B}) but not massive enough to fuse hydrogen in their cores are categorized as brown dwarfs. The dotted lines in Fig.~\ref{MIU} show clear clusters in each of the chosen celestial object range.

\subsection{Low-Mass Threshold}
 
To define the low-mass threshold of the FPP, we have chosen asteroids, moons, rocky exoplanets, and IAU planetary candidates\footnote{https://spaceflightnow.com/news/n0608/16planets/} for analysis. 
The data for 720 rocky exoplanets is taken from \cite{2023M}. The data for these objects is provided in Appendix~A (Tables 1 to 4). 

\begin{figure}
\centering 
\includegraphics[width=0.9\linewidth]{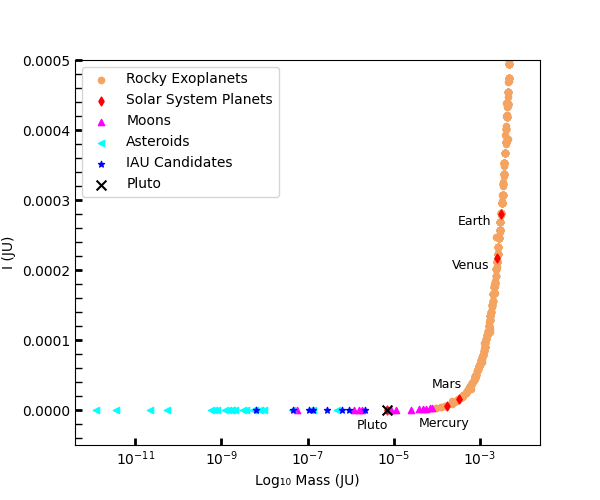}
\caption{Turn-off for rocky planets. The moment of inertia increases exponentially with mass for rocky planets, but is relatively constant for small-sized objects.}
\label{MIL}
\end{figure}

Using this data, we constructed Fig.~\ref{MIL}. Just as the stellar turn-off in the Hertzsprung-Russell diagram marks different stages in stellar evolution, we present a "planetary turn-off" diagram that includes rocky exoplanets, Solar System planets, major moons, IAU planetary candidates, asteroids, and Pluto. This representation reveals a clear distinction between planets and the other types of objects considered here.

The use of the term “turn-off” is motivated by analogy with the Hertzsprung–Russell diagram, where distinct regions correspond to physically meaningful stellar regimes. In the planetary context, the turn-off points identified in the Fundamental Planetary Plane should be interpreted as empirical boundaries that reflect transitions in dominant physical behavior, such as the onset of gravitational rounding, differentiation, or compressibility effects, rather than as evolutionary stages. These boundaries emerge naturally from the clustering of objects in parameter space and provide a practical means of distinguishing planets from other classes of celestial bodies.

In Fig.~\ref{MIL}, Mercury is used as a reference point -- being the smallest planet in the Solar System, it lies at the turn-off. Surrounding it are the smallest confirmed rocky exoplanets from the \cite{2023M} catalog, with the smallest one being PSR B1257+12b at 0.02 EU mass.\footnote{https://www.stellarcatalog.com}

Currently, it is impossible to observe or detect how clear the neighborhood of an exoplanet is. Hence, we propose the turn-off in the plot near Mercury and Mars in Fig.~\ref{MIL} as the low-mass threshold to define an object that is a planet. Also, we can confirm that the IAU planetary candidates are not planets according to this turn-off threshold.

\begin{figure}
\centering
\includegraphics[width=0.9\linewidth]{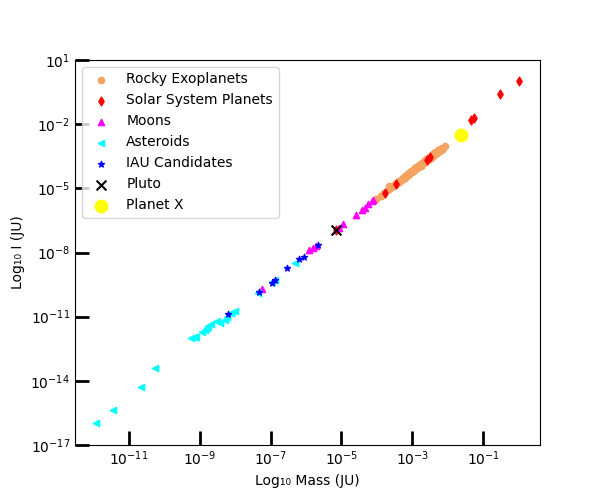}
\caption{Rocky Planet Plane: Red diamonds overlapping with brown dots which represent rocky planets in the Solar System. The black cross near the magenta triangle markers for moons corresponds to Pluto. Red diamonds located above the rocky exoplanets indicate giant planets of the Solar System. Blue stars represent IAU candidate objects under consideration for planetary status. Blue triangles denote asteroids.}
\label{log}
\end{figure}

If we re-plot Fig.~\ref{MIL} in the logarithmic scale, we can see a clustering trend at  different ranges (see Fig.~\ref{log}). These different planes for asteroids, moons, and planets can be defined as the fundamental asteroid plane, fundamental moon plane, and fundamental rocky planet plane, respectively. 


\subsection{Dwarf planets and Planet X (or Planet Nine)}

According to IAU definition, a "dwarf planet is a celestial body that (a) is in orbit around the Sun, (b) has sufficient mass for its self-gravity to overcome rigid body forces so that it assumes a hydrostatic equilibrium (nearly round) shape, (c) has not cleared the neighbourhood around its orbit, and 
(d) is not a satellite". The definition of a dwarf planet (like Pluto) is different from the brown dwarf (like Jupiter) definition. But dwarf galaxies are still galaxies, and similarly, dwarf stars are still stars (see discussion in Appendix~A). Therefore, dwarf planets can still be called planets. Ceres and Pluto are classified as dwarf planets because of the concept of internal differentiation. \cite{2022M} developed the argument that aligning the planetary taxonomy with geological complexity represents the most scientifically effective classification system. And they argued that moons are planets.

\begin{figure}
\centering    
\includegraphics[width=0.9\linewidth]{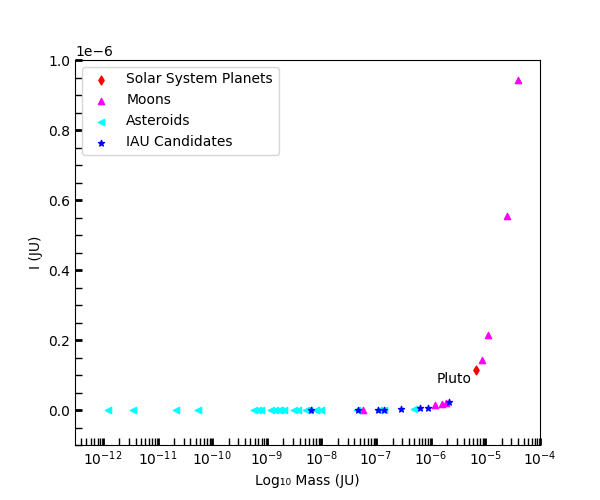}
\caption{Pluto turn-off in the moons plane: the magenta and blue colored triangles represent the moons and asteroids of the Solar System, respectively, while the blue stars denote IAU planetary candidates.}
\label{sub}
\end{figure}

Pluto's debate of being or not being a planet can be fixed using Fig~\ref{sub}. According to it, we confirm that Pluto is not a planet. We see a clear position of Pluto on the border of dwarf planets and asteroids. Pluto is shown as a fascinating world from the New Horizons 2015 flyby mission. Also, outer Solar System moons and some dwarf planets are promising objects to study for geology and even astrobiology.


\cite{2020N} results suggest that Ceres' material may have undergone aqueous alteration on small precursor bodies, with some remaining dry if the dwarf planet formed within Pluto’s orbit. Differentiation in the water percolation regime progresses gradually over several hundred million years, ultimately resulting in a partially differentiated structure. This includes a liquid layer that may still persist between the porous, undifferentiated crust and the silicate core, consistent with observations from the Dawn mission. Both Ceres and Pluto show internal differentiation, with a distinct core separate from the mantle.
Interestingly, there has been a new discovery recently of a dwarf planet called 2017 OF201 with a wide orbit (\citealt{2025C}).

A central pillar supporting the Planet Nine hypothesis lies in the unexpected clustering of orbital elements among distant trans-Neptunian objects (TNOs). Observations reveal that a subset of these objects -- particularly those with perihelia beyond 250 AU and semi-major axes exceeding 250 AU -- exhibit statistically significant alignment in both argument of perihelion and longitude of ascending node. This orbital confinement cannot be readily explained by interactions with the known planets, leading researchers to postulate the presence of an external gravitational perturber (\citealt{2019B}).

\subsection{High Mass Threshold}

\begin{figure}
\centering
\includegraphics[width=0.9\linewidth]{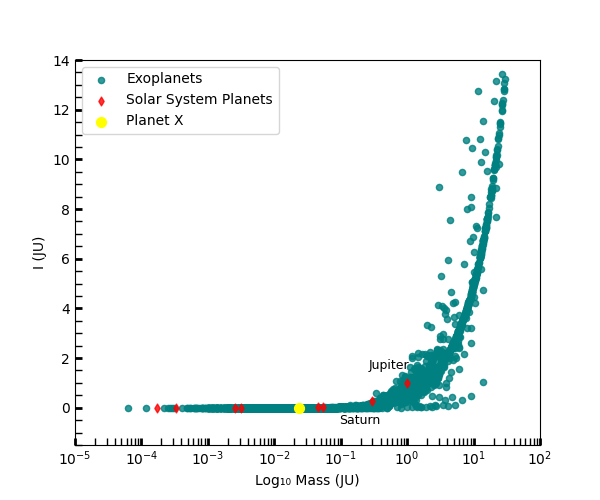}
\caption{Turn-off for gas giants: The gas giants of the Solar System can be identified by the red diamonds above the turn-off point, while the rocky planets and the ice giants lie below it.}
\label{MIgas}
\end{figure}

The upper mass limit threshold of the fundamental planetary plane may be more straightforward: there is a natural lower limit to what constitutes a star: $\sim$0.08 SU. 
But then there are brown dwarfs: IAU has defined brown dwarfs as objects that exceed the deuterium burning limit ($\sim$13 JU for solar metallicity) irrespective of formation mechanism. The distinction between brown dwarfs and actual stars is subtle: both low-mass stars and brown dwarfs burn primordial deuterium at first, but a star will eventually settle down to stable H-burning; in the case of a 0.08 SU star this will take $\sim$1 Gyr (\citealt{2001M}). 
Below a mass of 0.015 SU ($\sim$13 JU) not even burning of deuterium can occur, and these objects are perhaps best called planets. But it has been suggested that giant planets, formed by core accretion in a proplyd, can exceed the deuterium burning and, indeed, most exoplanets and brown dwarfs discovered to date include bodies ranging from $\sim$1 to $\sim$80 JU (e.g. \citealt{2015H}). In Fig.~\ref{MIgas}, we display a plot similar to Fig.~\ref{MIL} of Solar System planets and all exoplanets. Exoplanets data is from the Planetary Habitability Laboratory (PHL) database PHL-HWC\footnote{https://phl.upr.edu/hwc/data}. Here we see gas giants show a clear turn-off from the rocky planets. Similar to Fig.~\ref{MIL}, Saturn can be used as a reference point -- it lies at the turn-off (threshold), and ice giants are before this threshold. 

\begin{figure}
\centering
\includegraphics[width=0.9\linewidth]{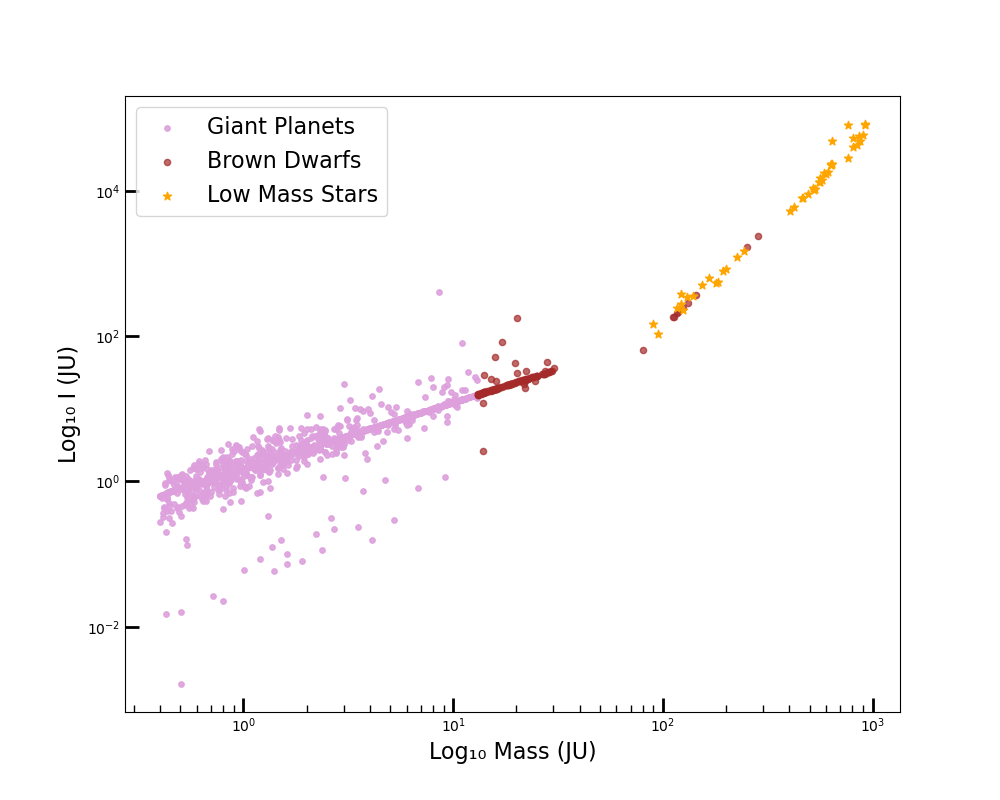}
\caption{Giant planets and low-mass stars: it is clearly seen that the slope of the low-mass stars plane is greater than the slope of the fundamental planes of giant planets and brown dwarfs, which indicates that hydrogen fusion in the core may play a role in increasing the moment of inertia of an object.}
\label{log2}
\end{figure}

When we re-plot Fig.~\ref{MIgas} in logarithmic scale, a clustering trend becomes evident across different ranges (see Fig.~\ref{log2}), corresponding to distinct planes for gas giants, brown dwarfs, and low-mass stars, respectively. The data for low-mass stars is taken from \citet{2017C}. We observe in Fig.~\ref{log2} a clear break between brown dwarfs and low-mass stars; this could be the radiation pressure threshold for hydrogen to fuse into helium. \\
Although the precise turnover in companion mass and mass-ratio distributions separating brown dwarfs from giant planets remains uncertain and requires further study, HIP~54515~b falls securely within the planetary regime for both metrics. Several directly imaged companions -- such as $\beta$~Pic~b--c, HR~8799~e, HIP~99770~b, $\epsilon$~Ind~b, and 14~Her~c -- exhibit comparable masses and mass ratios. Taken together, these comparisons indicate that HIP~54515~b is best classified as a super-Jovian planet rather than a brown dwarf \citep{Currie2025}.

Within the framework of the Fundamental Planetary Plane, Pluto and several IAU planetary candidates occupy regions distinct from those populated by confirmed planets. In particular, Pluto clusters with large moons and trans-Neptunian objects rather than with rocky planets. This result suggests that, according to the physical criteria adopted in this work, Pluto does not fall within the planetary regime. We emphasize that this conclusion is specific to the present framework, and alternative classification schemes based on different physical or geological criteria may lead to different interpretations.

\newpage
\section{The Mass–Radius Correlation: Artefact or Astrophysical Reality?}   

\begin{figure}
\centering
\includegraphics[width=0.9\linewidth]{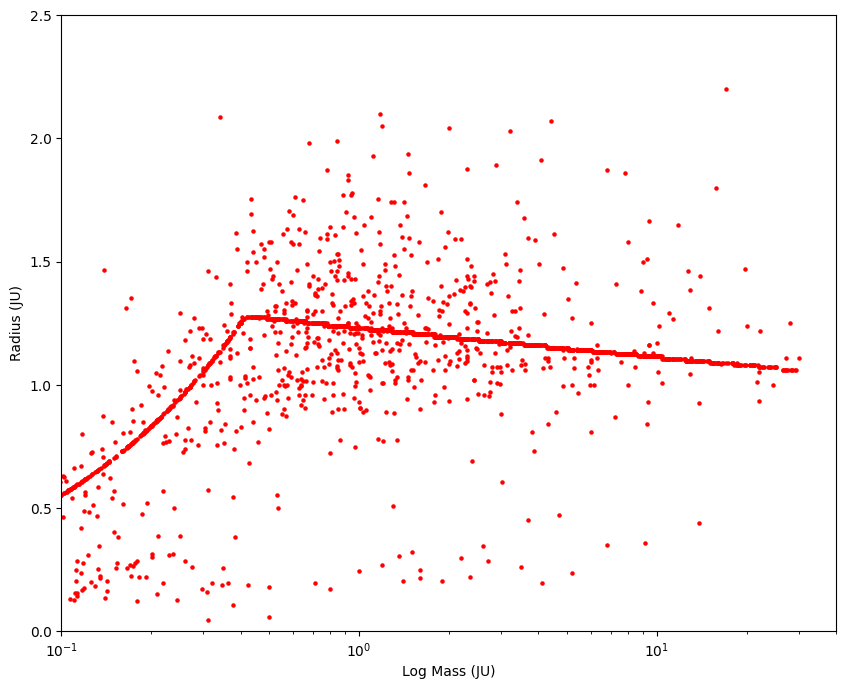}
\caption{Mass-Radius (MR) plot of exoplanets from PHL-HWC database (5592 exoplanets). The solid line is not a result of the fitting. We see an artefact which could be due to modeling.}
\label{MR}
\end{figure}

Studying the mass-radius relationship of exoplanets is crucial for understanding and characterizing their physical properties. All the exoplanetary system classifications depend on the mass-radius relationship (\citealt{2024P}). In Fig.~\ref{MR}, it seems that there is a fitted line, but it is an artifact. An artifact seen here arises naturally from the data. We checked this artifact over many other databases (such as The Exoplanet Data Explorer\footnote{http://exoplanets.org}, The Extrasolar Planets Encyclopaedia (https://exoplanets.eu), and NASA Exoplanet Archive (https://exoplanetarchive.ipac.caltech.edu/) in addition to the PHL–HWC (Planetary Habitable Laboratory - Habitable World Catalog). This could be due to the modeling. A similar pattern is also evident when examining the Kepler data alone.

\begin{figure}
\centering
\includegraphics[width=1\linewidth]{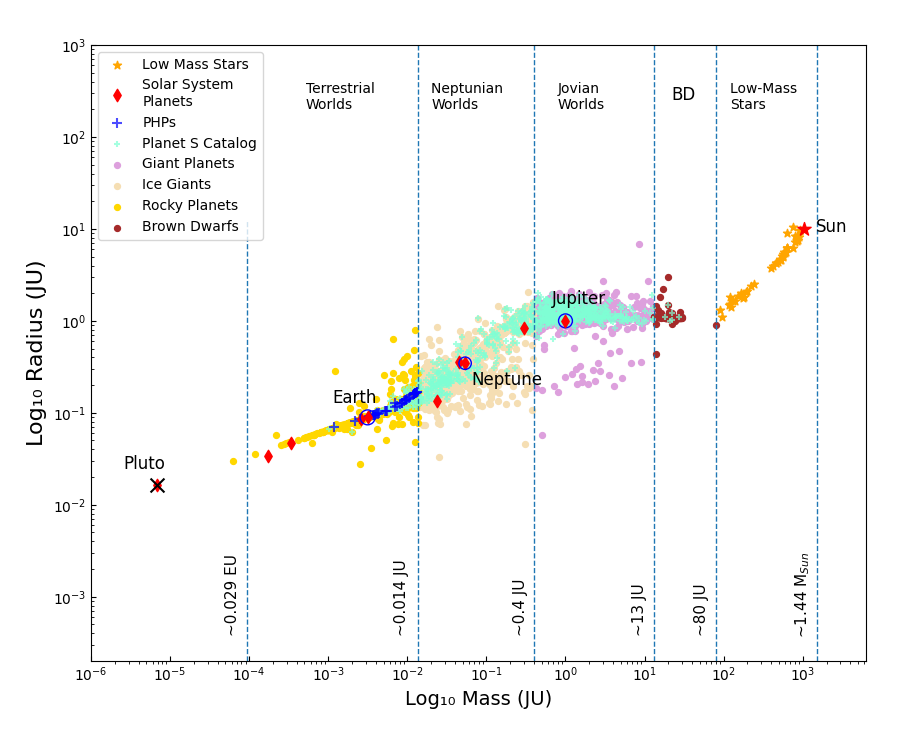}
\caption{
Position of 860 exoplanets with measured both $R$ and $M$ from the Planet S Catalog on the Fundamental Planetary Plane in comparison with Solar System planets, potentially habitable planets (PHPs), and low-mass stars. Vertical dashed lines indicate approximate regime boundaries. Representative Solar System bodies are labeled for reference.}
\label{MR2}
\end{figure}

To examine whether the observed Fundamental Plane represents a genuine physical trend or a modeling artefact (like in Fig.~\ref{MR}), 
we selected only the planets with both measured parameters from Planet S Catalog\footnote{Planet S Catalog: https://dace.unige.ch/exoplanets/}, and overlaid them on our FPP diagram (see Fig.~\ref{MR2}). The consistency of the measured data with the overall trend confirms that the observed relation is real rather than a modeling artefact. The Potentially Habitable Planets (PHPs) from the PHL database (Table~\ref{tab:placeholder_label}) are also overplotted to indicate the position of terrestrial worlds on the fundamental planetary plane.

\vspace{0.5cm}

\subsection{The radius-density corelation}\label{rd}

An alternative approach to identifying transitions among different classes of celestial bodies is to examine their distribution in the radius--density regime. For a spherical object, the bulk density is given by
\begin{equation}
    \rho = \frac{M}{\frac{4}{3}\pi R^{3}},
\end{equation}
where $M$ is the mass and $R$ is the radius. Because density reflects both internal composition and the degree of gravitational compression \citep{2007S, Fortney2007}, while radius traces the response of a body to increasing mass, the $\log\rho$--$R$ plane provides a physically informative diagnostic of planetary structure.

\begin{figure}
\centering
\includegraphics[width=1\linewidth]{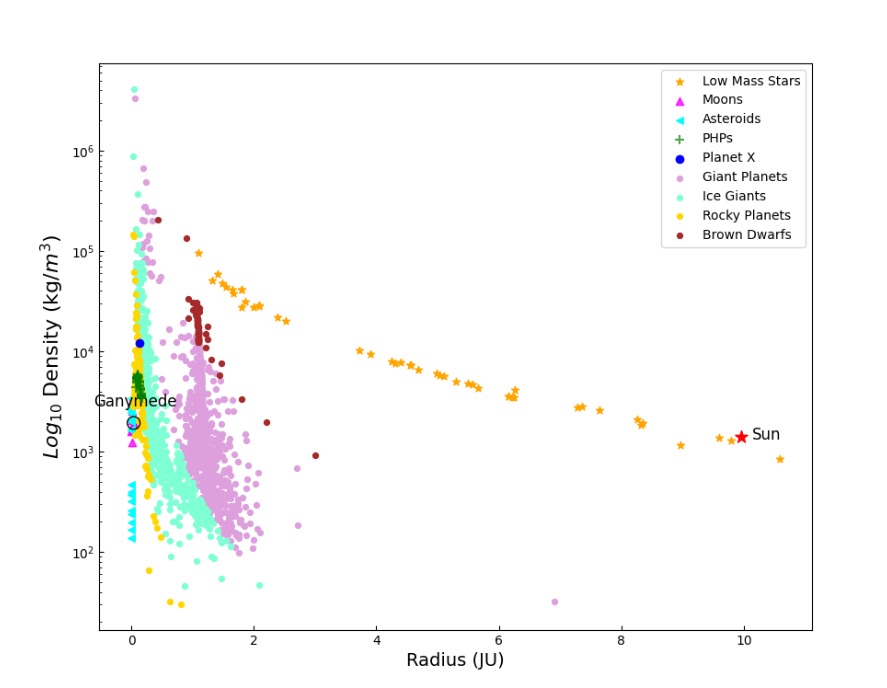}
\caption{
Logarithmic density–radius diagram (in JU units) for Solar System objects, exoplanets, brown dwarfs, and low-mass stars. The clustering reveals compositional and structural regimes, with rocky, icy, and gaseous planets distinct from the higher-radius, lower-density sub-stellar sequence.}
\label{RD}
\end{figure}
\vspace{0.5cm}

Figure~\ref{RD} shows the logarithmic density--radius distribution (in JU units) for Solar System bodies, exoplanets, brown dwarfs, and low-mass stars. Several trends are immediately apparent. Small irregular bodies -- including moons and asteroids -- occupy the low-radius, low-density region, consistent with porous interiors and the absence of significant self-gravity \citep{Carry2012,Fujiwara2006}. At higher densities and modest radii, rocky planets form a compact and well-defined cluster, characteristic of differentiated, gravitationally consolidated bodies \citep{McKinnon2015}. Beyond this regime, ice and gas giants extend toward progressively larger radii with correspondingly lower densities, reflecting their volatile-rich compositions \citep{Fortney2010}.

Brown dwarfs populate a distinct sequence where increasing mass leads to gravitational compression and only a limited change in radius, resulting in a systematic rise in density \citep{Chabrier2000}. Low-mass stars continue this trend toward even higher densities \citep{Baraffe1998}. This structural continuity across the sub-stellar sequence is clearly separated from the planetary domain.

Pluto’s position within this diagram is noteworthy: it lies within the cluster of large moons and trans-Neptunian objects rather than among the rocky planets. Its relatively low density and incomplete differentiation \citep{McKinnon2020} indicate that Pluto does not meet the structural criteria associated with true planets. The radius--density distribution therefore provides independent support, consistent with the fundamental planetary plane, that Pluto occupies the regime of minor bodies rather than planets.

Overall, the observed clustering and separations in the $\log\rho$--$R$ parameter space reinforce the physical transitions identified throughout this work, demonstrating that the radius--density plane naturally distinguishes between irregular bodies, planets, brown dwarfs, and low-mass stars.
 
\newpage
\section*{Summary}
We propose to define planets using the introduced Fundamental Planetary Plane, where, just like on the HR diagram for stars, the two parameters are correlated with the third.

\begin{itemize}   
\item   We observe a clear turn-off from the gas-giant planetary plane to low-mass stars (Fig.~\ref{MIU}), which we define as a giant planetary plane.
\item 
The HR diagram for stars shows a turn-off point of various stellar life cycles starting after star formation. Similarly, we see a turn-off for planets in Fig.~\ref{MIL} for rocky planets, which we define as the rocky planetary plane for low-mass planets.
\item 
In Fig.~\ref{sub}, we see that IAU planetary candidates are away from the fundamental plane of all known planetary objects. Hence, the IAU candidates are not planets, they are most probably asteroids.
\item 
According to Fig.~\ref{sub}, we confirm that Pluto is not a planet. We see a clear position of Pluto, which is not on the Fundamental Planetary Plane. Pluto is in the border between dwarf planets and asteroids. 
\item 
Fig.~\ref{MIgas} shows Solar System planets and exoplanets, where gas giants deviate from the rocky trend; Saturn marks the turn-off threshold, with ice giants lying below it.
\item
We notice an interesting break on a $\log M$-$\log I$ plane from brown dwarfs to low-mass stars in Fig.~\ref{log2}. This break could indicate the necessary threshold to begin the thermonuclear fusion reaction in stars.

\item 
The $\log\rho$–$R$ (Fig. \ref{RD}) plot naturally segregates minor bodies, planets, and sub-stellar objects, supporting the physical boundaries identified in the Fundamental Planetary Plane and confirming that Pluto clusters with moons rather than planets.

\item 
We re-define a planet as "A celestial spherical object, bound to a star or unbound, that lies on the Fundamental Planetary Plane, within a mass range between $\sim$$10^{23}$ Kg (0.02 EU) to $2.5 \times 10^{28}$ Kg (13 JU)".
The Fundamental Planetary Plane is intended as a structural classification framework that does not explicitly depend on formation mechanism. Nevertheless, planetary formation and thermal evolution are expected to influence internal structure and, consequently, the position of an object within the plane. Evolutionary processes may therefore shift objects within a given regime without necessarily moving them across the fundamental boundaries identified here. Future work may explore the extent to which evolutionary pathways can be mapped onto trajectories within the Fundamental Planetary Plane.

\end{itemize}

\section*{Acknowledgments}  

This research has made use of data provided by the portal exoplanet.eu of The Extrasolar Planets Encyclopaedia; the Exoplanet Orbit Database and the Exoplanet Data Explorer at exoplanets.org; The Habitable Worlds Catalog (HWC), PHL@UPR Arecibo; and NASA Exoplanet Archive, which is operated by the California Institute of Technology, under contract with the National Aeronautics and Space Administration under the Exoplanet Exploration Program.

\section*{Data Availability}   

All data generated or analyzed during this study are included in this published article.



\bibliographystyle{cas-model2-names}

\bibliography{cas-refs}

\section*{Appendix A. Tables with physical data}

\clearpage
\begin{table}[!b]
\centering
\caption{Solar System Planets (NASA planetary fact sheet)}
\label{table3}
\begin{tabular}{cccc}
\hline
\textbf{Name}&  \textbf{Mass (kg)} &\textbf{Mass (EU)}& \textbf{Radius (EU)}\\
\hline
         Mercury
&  $3.30\times10^{23}$ &$5.53\times10^{-2}$
& 0.38\\
         Venus
&  $4.86\times10^{24}$ &$0.81$
& 0.94\\
         Earth
&  $5.97\times10^{24}$ &1.00
& 1.00\\
         Mars
&  $6.39\times10^{23}$ &$0.10$
& 0.53\\
         Jupiter
&  $1.89\times10^{27}$ &$318.25$
& 11.21\\
         Saturn
&  $5.68\times10^{26}$ &95.30
& 9.45\\
         Uranus
&  $8.65\times10^{25}$ &14.50
& 4.01\\
 Neptune&  $1.02\times10^{26}$&17.10& 3.88\\
 Pluto& $1.30\times10^{22}$& $2.18\times10^{-3}$&0.18\\
 Planet X [{1}]& $4.47\times10^{25}$& 7.50&1.50\\
\hline
\multicolumn{4}{l}{[{1}] \cite{2019B}}\\
\hline
\end{tabular}
\end{table}

\begin{table}[h]
\centering
\caption{Largest Moons in the Solar System (NASA planetary fact sheet)}
\label{table4}
\begin{tabular}{cccc}
\hline
\textbf{Name}&  \textbf{Mass (kg)} &\textbf{Mass (EU)}& \textbf{Radius (EU)}\\
\hline
         Moon
&  $7.34\times10^{22}$ &$1.23\times10^{-2}$
& 0.27\\
         Io
&  $8.95\times10^{22}$ &$1.50\times10^{-2}$
& 0.28\\
         Europa
&  $4.80\times10^{22}$ &$8.04\times10^{-3}$
& 0.24\\
         Ganymede
&  $1.48\times10^{23}$ &$2.48\times10^{-2}$
& 0.41\\
         Callisto
&  $1.07\times10^{23}$ &$1.80\times10^{-2}$
& 0.37\\
         Enceladus
&  $1.07\times10^{20}$ &$1.80\times10^{-5}$
& 0.03\\
         Rhea
&  $2.30\times10^{21}$ &$3.86\times10^{-4}$
& 0.12\\
         Titan
&  $1.34\times10^{23}$ &$2.25\times10^{-2}$
& 0.40\\
 Titania
& $3.52\times10^{21}$ &$5.90\times10^{-4}$
&0.12\\
 Oberon
& $3.01\times10^{21}$ &$5.05\times10^{-4}$
&0.11\\
 Triton
& $2.13\times10^{22}$ &$3.58\times10^{-3}$
&0.21\\
 Eris
& $1.66\times10^{22}$ &$2.78\times10^{-3}$
&0.18\\
         Ceres
&  $9.55\times10^{20}$ &$1.60\times10^{-4}$& 0.07\\
\hline
\end{tabular}
\end{table}

\begin{table}
\centering
\caption{Largest Asteroids of the Solar System (NASA planetary fact sheet)}
\label{table1}
\begin{tabular}{cccc} 
\hline
\textbf{Name}&  \textbf{Mass (kg)} &\textbf{Mass (EU)}& \textbf{Radius (EU)}\\ 
\hline
         Ceres&  $9.39\times10^{20}$ &$1.57\times10^{-4}$
& 0.07\\ 
         Vesta&  $2.59\times10^{20}$ &$4.34\times10^{-5}$
& 0.04\\ 
         Pallas&  $2.11\times10^{20}$ &$3.53\times10^{-5}$
& 0.04\\ 
         Hygiea&  $8.32\times10^{19}$ &$1.39\times10^{-5}$
& 0.03\\ 
         Euphrosyne&  $1.87\times10^{19}$ &$3.13\times10^{-6}$
& 0.02\\ 
         Sylvia&  $1.48\times10^{19}$ &$2.48\times10^{-6}$
& 0.02\\ 
         Camilla&  $1.12\times10^{19}$ &$1.88\times10^{-6}$
& 0.02\\ 
         Hermione&  $1.00\times10^{19}$ &$1.68\times10^{-6}$
& 0.01\\ 
 Davida& $5.98\times10^{18}$ &$1.00\times10^{-6}$
&0.02\\ 
 Elektra& $7.00\times10^{18}$ &$1.17\times10^{-6}$
&0.01\\ 
 Interamnia& $3.90\times10^{18}$ &$6.53\times10^{-7}$
&0.02\\ 
 Europa& $3.27\times10^{18}$ &$5.48\times10^{-7}$
&0.02\\ 
 Eunomia& $3.12\times10^{18}$ &$5.23\times10^{-7}$
&0.02\\ 
 Juno& $2.67\times10^{18}$ &$4.47\times10^{-7}$
&0.02\\ 
 Psyche&$ 2.27\times10^{18}$ &$3.80\times10^{-7}$
&0.02\\ 
 Thisbe& $1.46\times10^{18}$ &$2.45\times10^{-7}$
&0.02\\ 
 Bamberga& $1.28\times10^{18}$ &$2.14\times10^{-7}$
&0.02\\ 
 Cybele& $1.08\times10^{18}$ &$1.81\times10^{-7}$
&0.02\\ 
 Mathilde& $1.03\times10^{17}$ &$1.73\times10^{-8}$
&$8.27\times10^{-3}$\\ 
 Ida& $4.20\times10^{16}$ &$7.04\times10^{-9}$
&$2.46\times10^{-3}$\\ 
 Eros& $6.69\times10^{15}$ &$1.12\times10^{-9}$
&$1.31\times10^{-3}$\\ 
 Gaspra& $2.30\times10^{15}$ &$3.85\times10^{-10}$&$9.56\times10^{-4}$\\ 
\hline
\end{tabular}
\end{table}

\begin{table}
\caption{IAU Planetary Candidates Watchlist \\}
\label{table2}
\begin{tabular}{cccc}
\hline
\textbf{Name}&  \textbf{Mass (kg)} &\textbf{Mass (EU)}& \textbf{Radius (EU)} \\
\hline
Haumea&  $4.00\times10^{21}$ &$6.71\times10^{-4}$ [\textcolor{blue}{1}] 
         & 0.11 [\textcolor{blue}{2}]  \\
         Sedna
&  $1.70\times10^{21}$ &$2.85\times10^{-4}$ [\textcolor{blue}{3}]  & 0.07 [\textcolor{blue}{3}] \\
         Orcus
&  $5.47\times10^{20}$ &$9.16\times10^{-5}$ [{\textcolor{blue}{5}}]& 0.07 [{\textcolor{blue}{6}}] \\
         Quaoar
&  $1.20\times10^{21}$
 &$2.01\times10^{-4}$ [{\textcolor{blue}{7}}]& 0.08 [{\textcolor{blue}{7}}] \\
         2002 TX 300
&  $1.20\times10^{19}$ &$2.01\times10^{-6}$ [{\textcolor{blue}{8}}]& 0.02 [{\textcolor{blue}{8}}] \\
         Vesta
& $2.59\times10^{20}$ &$4.33\times10^{-5}$ [{\textcolor{blue}{12}}]&0.04 [{\textcolor{blue}{12}}] \\
 Pallas
& $2.04\times10^{20}$ &$3.41\times10^{-5}$ [{\textcolor{blue}{13}}]&0.04 [{\textcolor{blue}{13}}] \\
 Hygiea& $8.74\times10^{19}$ &$1.46\times10^{-5}$ [{\textcolor{blue}{14}}]&0.03 [{\textcolor{blue}{14}}] \\
\hline
References to the data: \\
\multicolumn{4}{|l|}{[{1}] \cite{HamueaM}, [{2}] \cite{HamueaR},} \\
\multicolumn{4}{|l|}{[{3}] \cite{SednaR},}\\ 
\multicolumn{4}{|l|}{[{5}] \cite{OrcusM}, [{6}] \cite{OrcusR},} \\
\multicolumn{4}{|l|}{[{7}] \cite{QuaoarR}, [{8}] \cite{tx300M}, 
\cite{VestaR},} \\
\multicolumn{4}{|l|}{[{13}] \cite{Pallas}, [{14}] \cite{Hygiea}} \\
\hline
\end{tabular}
\end{table}

\newpage
\clearpage

\begin{table}[t!]
    \centering
    \small
    \caption{Potential Habitable Planets from PHL database}
    \label{tab:placeholder_label}
    \begin{tabular}{cccc}
        \hline
        \textbf{Name}& \textbf{Mass (kg)}& \textbf{Mass (EU)}& \textbf{Radius (EU)}\\
        \hline
        TOI-700 e
& $4.84\times10^{24}$  
& 0.81
& 0.95
\\
        TOI-700 d
& $7.46\times10^{24}$  
& 1.25
& 1.07
\\
        Teegarden's Star b
& $6.93\times10^{24}$  
& 1.16
& 1.05
\\
        Teegarden's Star c
& $6.27\times10^{24}$  
& 1.05
& 1.02
\\
        Kepler-452 b
& $1.96\times10^{25}$  
& 3.29
& 1.63
\\
 Kepler-1649 c
& $7.16\times10^{24}$  
& 1.20&1.06
\\
 Kepler-1544 b
& $2.28\times10^{25}$  
& 3.82
&1.78
\\
 Kepler-1410 b
& $2.28\times10^{25}$  
& 3.82
&1.78
\\
 Kepler-296 f
& $2.32\times10^{25}$  
& 3.89
&1.80\\
 TRAPPIST-1 g
& $7.88\times10^{24}$  
& 1.32
&1.12
\\
 Kepler-283 c
& $2.37\times10^{25}$  
& 3.97
&1.82
\\
 TRAPPIST-1 d
& $2.27\times10^{24}$  
& 0.38
&0.78
\\
 GJ 1002 b
& $6.45\times10^{24}$  
& 1.08
&1.03
\\
 Wolf 1061 c
& $2.04\times10^{25}$  
& 3.41
&1.66
\\
 GJ 667 C e
& $1.61\times10^{25}$  
& 2.70&1.45
\\
 GJ 667 C f
& $1.61\times10^{25}$  
& 2.70&1.45
\\
 TRAPPIST-1 e
& $4.12\times10^{24}$  
& 0.69
&0.92
\\
 TOI-715 b
& $1.80\times10^{25}$  
& 3.02
&1.55
\\
        GJ 667 C c
& $2.27\times10^{25}$  
& 3.80& 1.77
\\
        K2-288 B b
& $2.55\times10^{25}$  
& 4.27
& 1.90\\
        
        Proxima Cen b
& $6.39\times10^{24}$  
& 1.07
& 1.03
\\

        K2-3 d
& $1.31\times10^{25}$  
& 2.20& 1.45
\\
        Kepler-440 b
& $2.46\times10^{25}$  
& 4.12
& 1.86
\\
        Kepler-442 b
& $1.41\times10^{25}$  
& 2.36
& 1.34
\\
        Ross 128 b
& $8.36\times10^{24}$  
& 1.40& 1.11
\\
        GJ 273 b
& $1.73\times10^{25}$  
& 2.89
& 1.51
\\
        TRAPPIST-1 f
& $6.15\times10^{24}$  
& 1.03
& 1.04
\\
        GJ 1061 c
& $1.04\times10^{25}$  
& 1.74
& 1.18
\\
        Kepler-1638 b
& $2.48\times10^{25}$  
& 4.16
& 1.87
\\
        Kepler-1652 b
& $1.90\times10^{25}$  
& 3.19
& 1.60\\
        Kepler-186 f
& $1.02\times10^{25}$  
& 1.71
& 1.17
\\
        Kepler-1229 b
& $1.52\times10^{25}$  
& 2.54
& 1.40\\
        GJ 1002 c
& $8.12\times10^{24}$  
& 1.36
& 1.10\\
        GJ 1061 d
& $9.79\times10^{24}$  
& 1.64
& 1.16
\\
        Kepler-296 e
& $1.77\times10^{25}$  
& 2.96
& 1.53
\\
        K2-72 e
& $1.32\times10^{25}$  
& 2.21
& 1.29
\\  
        Ross 508 b
& $2.39\times10^{25}$  
& 4.00& 1.83
\\
        Wolf 1069 b& $7.52\times10^{24}$& 1.26& 1.08\\
        \hline 
    \end{tabular}
\end{table}

\end{document}